\documentclass[aps,prl,twocolumn,showpacs]{revtex4}
\usepackage{graphics} 
\usepackage{amssymb}
\usepackage{amsmath}
\usepackage{bm}

\usepackage[pdftex]{graphicx}

\begin{document}

\title{Observation of a Geometric Hall Effect in a Spinor Bose-Einstein\\ Condensate with a Skyrmion Spin Texture}
\author{Jae-yoon Choi, Seji Kang, Sang Won Seo, Woo Jin Kwon, and Yong-il Shin}
\email{yishin@snu.ac.kr}

\affiliation{Center for Subwavelength Optics and Department of Physics and Astronomy, Seoul National University, Seoul 151-747, Korea}

\begin{abstract}
For a spin-carrying particle moving in a spatially varying magnetic field, effective electromagnetic forces can arise due to the geometric phase associated with adiabatic spin rotation of the particle. We report the observation of a geometric Hall effect in a spinor Bose-Einstein condensate with a skyrmion spin texture. Under translational oscillations of the spin texture, the condensate resonantly develops a circular motion in a harmonic trap, demonstrating the existence of an effective Lorentz force. When the condensate circulates, quantized vortices are nucleated in the boundary region of the condensate and the vortex number increases over 100 without significant heating. We attribute the vortex nucleation to the shearing effect of the effective Lorentz force from the inhomogeneous effective magnetic field.
\end{abstract}


\maketitle

When a particle with a spin slowly moves in a spatially varying magnetic field and its spin adiabatically follows the field direction, the particle acquires a quantum-mechanical phase known as the Berry phase~\cite{Berry_ProcRoyal}. This phase originates from the geometrical properties of the parameter space of the system and in the Hamiltonian description, it can be represented as a gauge potential~\cite{GeometricW}. Just like the vector potential for a charged particle, the gauge potential generates forces from its spatial and temporal variations~\cite{Stern}, and the geometric forces act like magnetic and electric forces on the spin-carrying particle.

Emergent electromagnetism of this spin origin can lead to novel spin transport phenomena. In magnetic materials, because of the coupling of a spin current to magnetization, a non-coplanar spin texture of magnetization gives rise to an effective internal magnetic field, leading to the intrinsic anomalous Hall effect~\cite{AHE_science,AHE_RMP}. Recently, the topological Hall effects were observed in chiral magnets with skyrmion lattice spin textures~\cite{THE_PRL1,THE_PRL2,THE_PRL3}. Since the Berry phase is proportional to the spin value, effective electromagnetic forces are intrinsically spin dependent and, thus, they have been actively investigated for spintronics applications~\cite{SMF_PRL,SMF_Nature,SMF_DW}. In recent experiments with ultracold neutral atoms, artificial magnetic and electric fields were synthesized using atom-light interaction~\cite{Lightgauge_Nature,eforce_Natphys}, presenting a new opportunity for exploring quantum many-body phenomena in gauge fields. This experimental technique is based on the Berry-phase effect from the pseudo-spin texture of light-dressed atoms~\cite{Dalibard_review}.

In this Letter, we report the observation of a geometric Hall effect in a spin-polarized atomic Bose-Einstein condensate, where a rigid spin texture of skyrmion configuration is imposed by a spatially varying external magnetic field. We investigate the condensate dynamics under translational oscillations of the spin texture and observe that a circular motion of the condensate is resonantly induced from the translational drive. This directly manifests the existence of the effective Lorentz force acting on the condensate. Furthermore, we observe that quantized vortices are dynamically nucleated in the circulating condensate, which we attribute to the inhomogeneity of the effective magnetic field. 

This work presents an alternative method for generating an artificial gauge field for neutral atoms, in particular, without using light. Based on spatial and temporal control of the real atomic-spin texture, this method does not suffer from atom loss and heating associated with spontaneous scattering of light~\cite{Dalibard_review,SOreview}, and thus enables us to study superfluid dynamics in a gauge field in a more dissipation-free condition~\cite{SuperfluidHall}.

We consider a neutral atom with hyperfine spin $F$ moving in a spatially and temporally varying magnetic field $\mathbf{B}(\mathbf{r},t)$. The system's Hamiltonian is given as $H=\mathbf{p}^2/2m -  g_F\mu_B \mathbf{F}\cdot \mathbf{B}$, where $m$ is the atomic mass, $g_F$ is the Land\'{e} $g$ factor, $\mu_B$ is the Bohr magneton, and $\mathbf{F}$ is the spin operator. If we take a local unitary transformation $U=\exp (-i\beta\hat{n}\cdot\mathbf{F})$ that rotates the quantization axis for the spin from a fixed axis $\hat{z}$ to the local magnetic field direction $\hat{b}=\mathbf{B}/|\mathbf{B}|$, where $\beta$ is the tilting angle of $\hat{b}$ from $\hat{z}$ and $\hat{n}=(\hat{b}\times\hat{z})/|\hat{b}\times\hat{z}|$, then the Hamiltonian is transformed to $H'=(\mathbf{p}+\hbar\mathcal{A})^2/2m+V-g_F\mu_B F_z |\mathbf{B}|$ with a gauge potential $\mathcal{A}(\mathbf{r},t)=-i U \nabla U^\dag$ and a scalar potential $V(\mathbf{r},t)=-i\hbar U \partial_t U^\dag$, where $\hbar$ is the Planck constant $h$ divided by 2$\pi$. In the adiabatic regime where the atom keeps its spin in the $m_F$ state for the quantization axis along $\hat{b}(\mathbf{r},t)$, the atom feels effective magnetic and electric fields~\cite{Hogauge_PRL,SpintronicsReview}:
\begin{eqnarray}
\mathbf{B}^{e}_i &=& -m_F \frac{\hbar}{2} \epsilon_{ijk}\hat{b}\cdot(\partial_j\hat{b}\times\partial_k\hat{b}) \\ 
\mathbf{E}^{e}_i &=& -m_F \hbar \hat{b}\cdot(\partial_i\hat{b}\times\partial_t\hat{b}),
\end{eqnarray}
where $i,j,k$ are spatial coordinates and $\epsilon_{ijk}$ is the Levi-Civita symbol. The corresponding charge value of the atom is set to be unity.

\begin{figure}
\includegraphics[width=8.3cm]{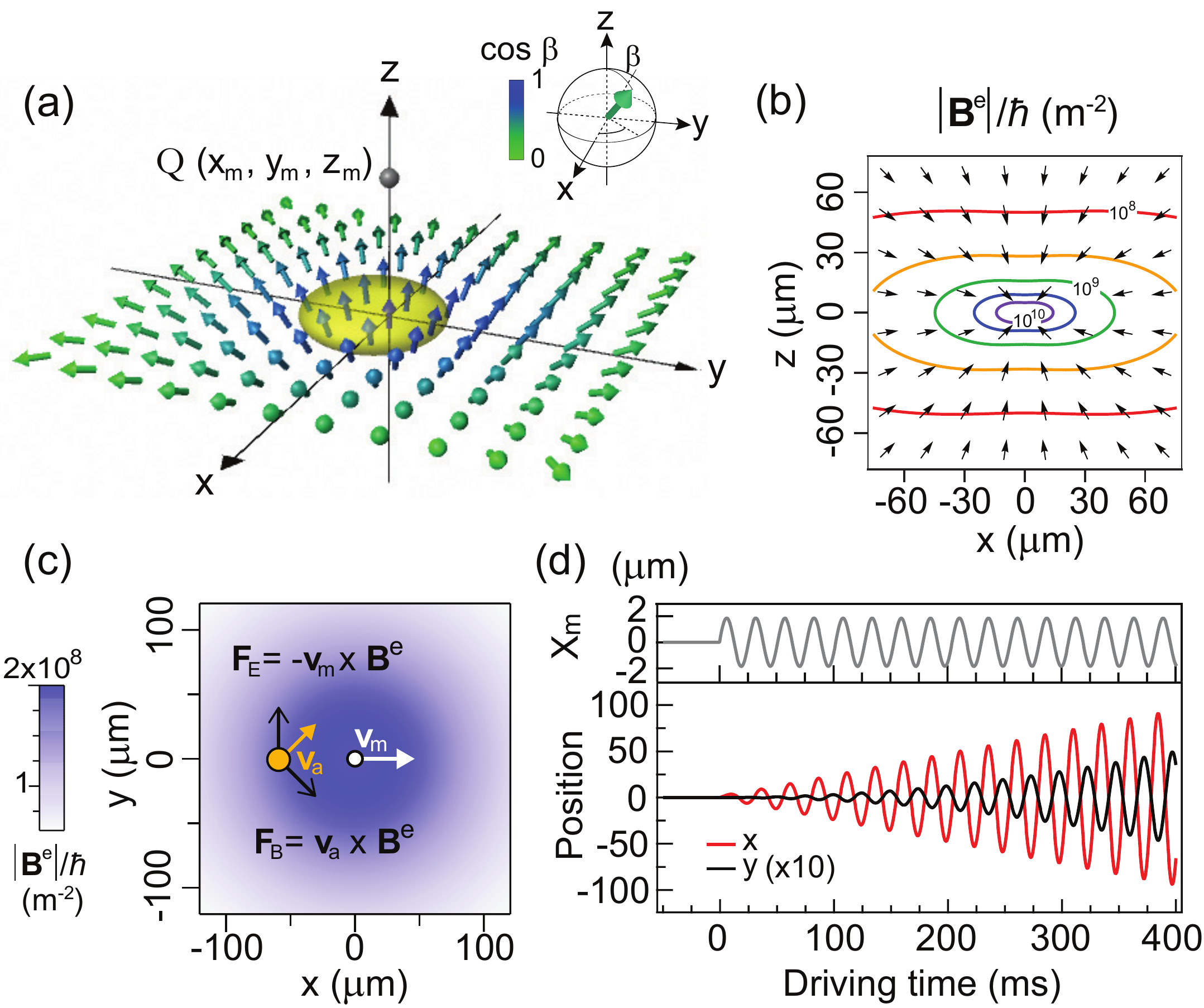}
\caption{(color online). Geometric Hall effect for a neutral atom moving in a three-dimensional magnetic quadrupole field. (a) The zero-field center of the magnetic field is indicated by a point $Q$. The atom (yellow ellipsoid) is in the $|m_F=-1\rangle$ spin state for the local field direction denoted by an arrow. (b) Effective magnetic field $\mathbf{B}^e$ for the atom. (c) When the atom and the zero-field center move with velocities $\mathbf{v}_a$ and $\mathbf{v}_m$, respectively, an effective Lorentz force $\mathbf{F}_B=\mathbf{v}_a\times\mathbf{B}^e$ and an effective electric force $\mathbf{F}_E=-\mathbf{v}_m\times\mathbf{B}^e$ occur on the atom. (d) The Hall response of the atom to sinusoidal oscillations of the zero-field center [Eq.~(6) for $\omega_m=\omega_r$ and $X_m=2~\mu$m]. A $y$-directional motion emerges from the $x$-directional modulation of $Q$.}
\label{Figure1}
\end{figure}

Our system consists of a Bose-Einstein condensate of $^{23}$Na atoms in the $|F=1,m_F=-1\rangle$ state. We prepare a nearly pure condensate in a pancake-shaped optical dipole trap with trapping frequencies of $\omega_{x,y,z}=2\pi\times (3,3.9,370)$~Hz and adiabatically ramp-up a three-dimensional magnetic quadrupole field
\begin{equation}
\mathbf{B}=\frac{B_q}{2}(x\hat{x}+y\hat{y}-2z\hat{z}) - B_x\hat{x}+B_z\hat{z}
\end{equation}
with $B_q=7.6$~G/cm. The position of the zero-field center is controlled by external bias fields $B_x$ and $B_z$ as $\mathbf{r}_m=(x_m,y_m,z_m)=(\frac{2B_x}{B_q},0,\frac{B_z}{B_q})$ and initially placed above the condensate at $\mathbf{r}_m=(0,0,36)~\mu$m~[Fig.~1(a)]. In the $z=0$ plane, the tilt angle $\beta$ of $\hat{b}$ increases with radial position $r'_{\perp}=\sqrt{(x-x_m)^2+(y-y_m)^2}$ as $\tan \beta=\frac{r'_\perp}{2z_m}$ and the magnetic field imposes a spin texture of two-dimensional skyrmion configuration on the condensate~\cite{SNUskyr_PRL}. The Zeeman energy $E_\textrm{Z}=-g_F \mu_B |\vec{B}|$, with $g_F=-\frac{1}{2}$, provides an additional but dominant radial trapping potential. The radial trapping frequency is estimated to be $\omega_r=\sqrt{\omega_{x,y}^2+\frac{\mu_B}{8m}\frac{B_q}{z_m}}\approx 2\pi\times 40.3$~Hz at the trap center. For the typical atom number of $N\approx 3.3\times 10^6$, the Thomas-Fermi radius of the condensate is $R_\textrm{TF}\approx 44~\mu$m.

\begin{figure}
\includegraphics[width=8.5cm]{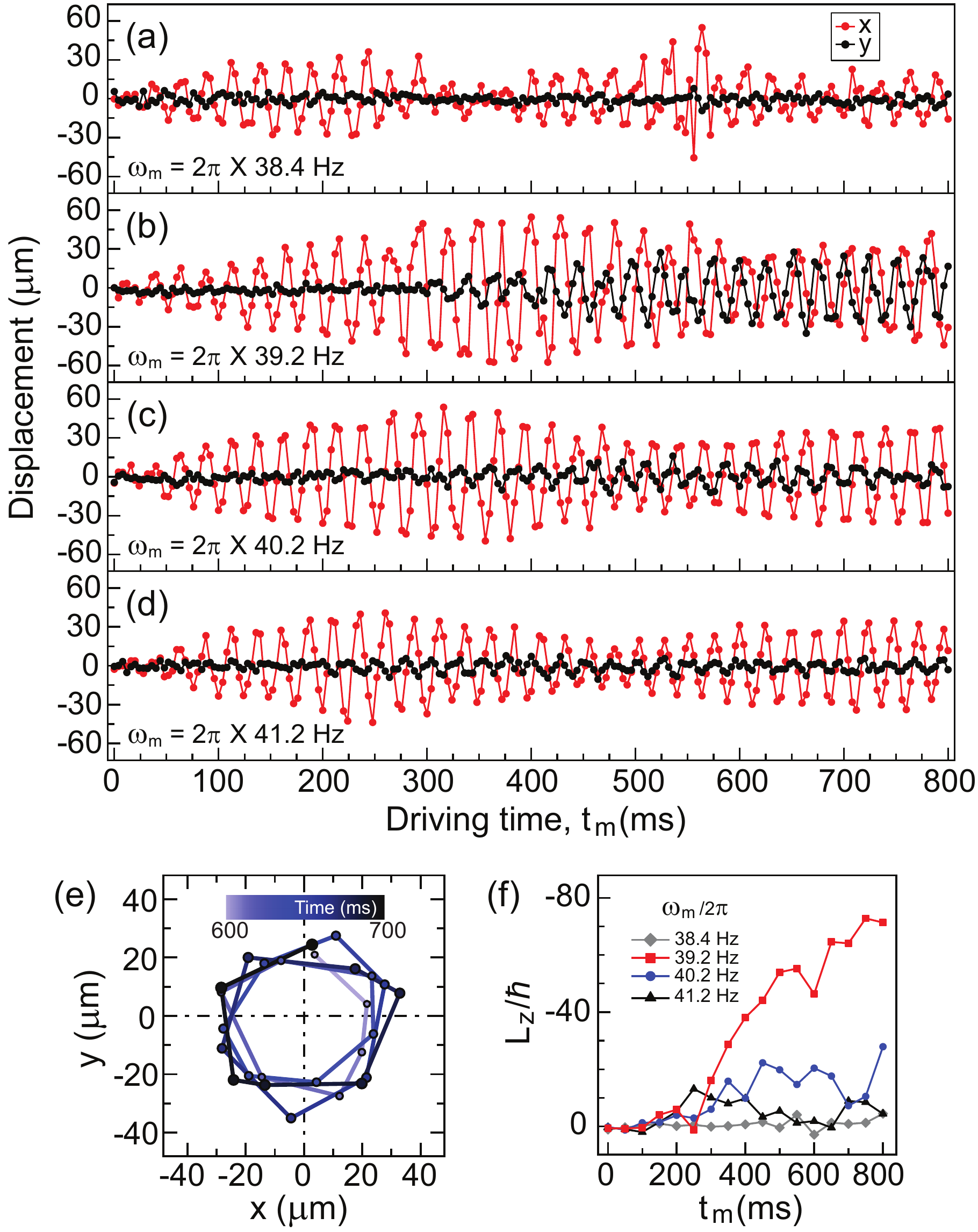}
\caption{(color online). (a)-(d) Temporal evolution of the condensate position under translational oscillations of the spin texture for various driving frequencies $\omega_m$. The condensate position was measured for every 4~ms and each data point was obtained from a single measurement. (e) Trajectory of the condensate in the $x$-$y$ plane for $\omega_m=2\pi\times39.2$~Hz. (f) External angular momentum $L_z$ per atom in the condensate. The condensate velocity was determined from the position data and $L_z$ is displayed with 50 ms binning.}
\label{Figure2}
\end{figure}

For the spin-polarized condensate in the $|m_F=-1\rangle$ state, the gauge potential and the effective magnetic field are given as
\begin{eqnarray}
\mathcal{A}(\mathbf{r})&=& - \frac{ \sqrt{r'^2_{\perp}+4z'^2} +2z' }{r'_{\perp}\sqrt{r'^2_{\perp}+4z'^2}}~\hat{\theta}
\\
\mathbf{B}^e(\mathbf{r})&=&-\frac{2\hbar}{ (r'^2_{\perp}+4z'^2)^{3/2}}~\mathbf{r}',
\end{eqnarray}
where $\hat{\theta}$ is the unit vector of the azimuthal direction with respect to the $+z$ axis and $\mathbf{r}'=\mathbf{r}-\mathbf{r}_m$~\cite{SNUGauge_JKPS}. We can associate the effective magnetic field $\mathbf{B}^e$ with a monopole located at $\mathbf{r}_m$ which carries effective flux of $\oint \mathbf{B}^e \cdot d\mathbf{\sigma}= -2 h$, ~[Fig.~1(b)]~\cite{Mono_PRL,unwinding}. The magnetic flux passing through the condensate is $\Phi_{B}^e/h=0.13$, which is much smaller than the critical value for having a singular vortex ground state~\cite{Hogauge_PRL,Vcriti_PRA,SNUGauge_JKPS}.

\begin{figure*}
\includegraphics[width=17cm]{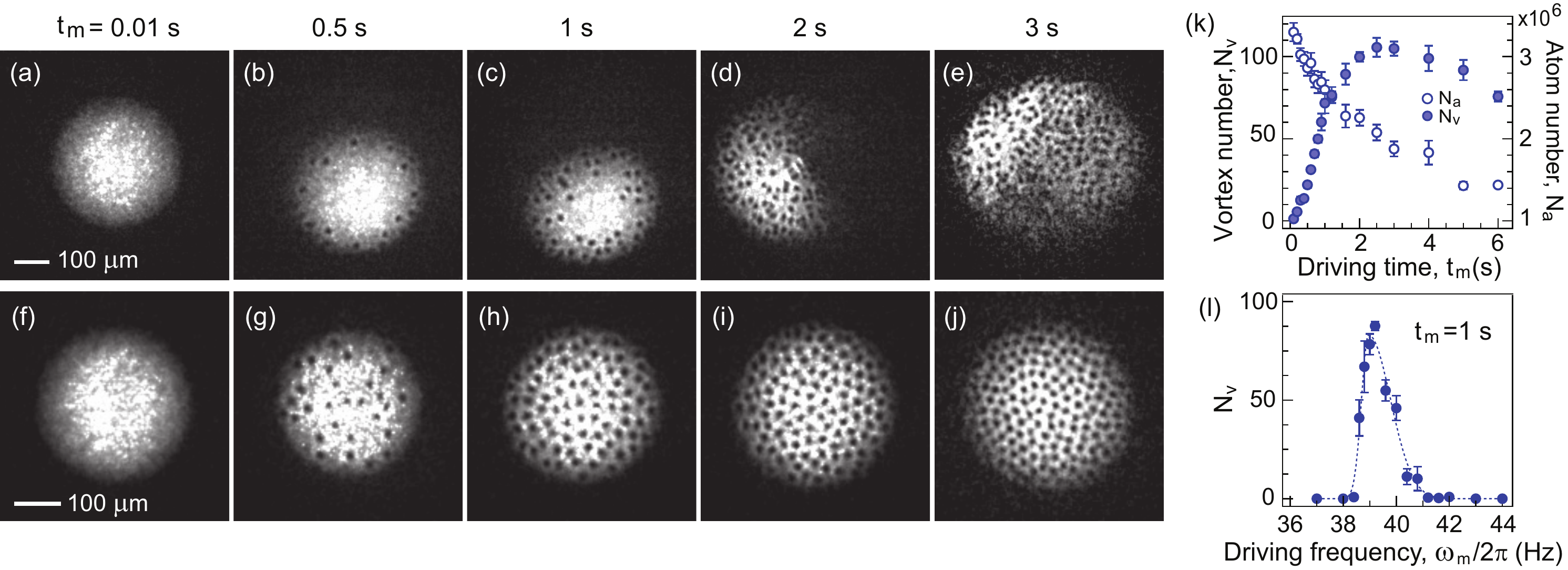}
\caption{(color online). Vortex nucleation in the spinor condensate circulating around the skyrmion spin texture. (a)-(e) Absorption images after 15-ms time-of-flight for $\omega_m=2\pi\times39.2$~Hz and various driving times $t_m$. The condensate deforms into an elliptical shape and quantized vortices are nucleated in its boundary. (f)-(j) Images after 2~s relaxation in a stationary trap~\cite{Supple}. (k) Temporal evolution of the vortex number $N_v$ (solid circles) and the atom number $N_a$ (open circles). (l) $N_v$ for $t_m=1$~s versus $\omega_m/2\pi$. The dashed line is an asymmetric Gaussian curve fit to the data, giving the center frequency at 39.2~Hz, and the left and right $1/e^2$ linewidths of 0.4 and 1.2~Hz, respectively. Each data point was obtained from at least eight measurements.}
\label{Figure3}
\end{figure*}

We study a Hall response of the system under translational oscillations of the spin texture, $\mathbf{r}_m(t)= X_{m}\sin \omega_m t~\hat{x}$, which is driven by sinusoidal modulation of the external bias field $B_x$.  The temporal variation of the spin texture generates an effective electric field $\mathbf{E}^e=-\dot{\mathbf{r}}_m \times \mathbf{B}^e$ [Eq.~(2)], perpendicular to the driving direction $\hat{x}$. This field can be understood as an induced electric field from the moving magnetic monopole with velocity $\mathbf{v}_m=\dot{\mathbf{r}}_m$ [Fig.~1(c)]. Including the restoring force from the magnetic trapping potential, the equation of motion for a single atom in the $z=0$ plane is given as
\begin{equation}
 m\frac{d^2 \mathbf{r}'_\perp}{dt^2}  = -m\omega_{r}^2\mathbf{r}'_\perp + \frac{d\mathbf{r}'_\perp}{dt}\times {B}^e_z(r_{\perp}')\hat{z}+ m\omega_{m}^2 X_m \sin\omega_{m}t~\hat{x},
\end{equation}
where $\mathbf{r}'_\perp=(x-x_m,y)$ and ${B}^e_z(r_{\perp}')=\frac{2\hbar z_m}{(r_\perp'^2+4z_m^2)^{3/2}}$ from Eq.~(5)~\cite{Supple}. In the reference frame of the spin texture, this describes a driven harmonic oscillator under a static magnetic field. When we start with an atom at rest, the atom will undergo forced oscillations along the $x$ direction and its transverse motion will be induced by the effective Lorentz force [Fig.~1(d)]. This is a geometric Hall effect arising from the non-coplanar spin texture.

In our experimental condition, the cyclotron frequency $\omega_c=\frac{|\mathbf{B}^e|}{m}\leq\frac{\hbar}{4m z_m^2}$ is 3 orders of magnitude smaller than the trap frequency $\omega_r$, implying that it would be experimentally challenging to directly detect the Lorentz force by examining an atom's motion. However, the effects of the effective Lorentz force can be amplified through resonant behavior of the system because the trapping frequencies for the $x$ and $y$ directions are same. When $\omega_m=\omega_r$ in Eq.~(6), the $x$-directional amplitude linearly increases as $\frac{\omega_r X_m}{2}t$, leading to a quadratic amplification of the $y$-directional motion as $\frac{\omega_c \omega_r X_m}{8 m} t^2$. For $\omega_c/\omega_r\sim 10^{-3}$, the $y$-directional amplitude is expected to become comparable to the modulation amplitude $X_m$ within 0.5~s [Fig.~1(d)].

We investigate the center-of-mass motion of the condensate by taking its \textit{in situ} absorption images for various driving times $t_m$ and $X_m\approx2~\mu$m~\cite{footnote2}. The condensate position was determined from a 2D Gaussian fit to the images. Figures~2(a)-2(d) display the temporal evolutions of the displacement of the condensate from its initial position for various driving frequencies $\omega_m$ near the resonance. The initial increasing rate of the $x$-directional amplitude is measured to be about 2.3$\times10^{-4}$~m/s, which is close to the estimated value of $\frac{\omega_r X_m}{2}=2.5\times 10^{-4}$~m/s, and discernible $y$-directional motion appears after a few 100~ms. In the peak response at $\omega_m=2\pi\times 39.2$~Hz, we observe that a circular motion with almost zero ellipticity develops for $t_m>600$~ms [Fig.~2(e)]. The radius of the trajectory is about 30~$\mu$m, corresponding to external angular momentum $L_z\approx-80\hbar$ per atom. The emergence of a circular motion under the translational drive characterizes the geometric Hall effect in our system.

We observe that the circulation direction of the condensate is always clockwise, giving negative $L_z$~[Fig.~2(f)]. Numerical simulation of the single-particle dynamics predicts circulation in the opposite direction for large negative detuning of the driving frequency, $(\omega_m-\omega_r)/2\pi<-1$~Hz~\cite{Supple}. However, in our experiment for $\omega_m<2\pi\times 38.5$~Hz, there is no detectable $y$-directional motion and even the development of $x$-directional oscillation seems to be strongly suppressed [Fig.~2(a)]. Full accounting for the condensate dynamics requires a superfluid hydrodynamics description including the interplay between the spin texture and the internal mass current~\cite{Lama_PRA,Ueda_PRep}. In the gauge field, the superfluid velocity $\mathbf{v}_s$ is required to satisfy the Mermin-Ho relation $\nabla\times \mathbf{v}_s=-\mathbf{B}^e/m$~\cite{Hogauge_PRL}, and therefore the condensate always has an internal rotational motion. Note that the initial state of the condensate is a coreless vortex state with $\mathbf{v}_s=\frac{\hbar}{m}\mathcal{A}$ due to the skyrmion spin texture~\cite{CorelessV,AngQT,Choi_NJP}. We speculate that the external circulation of the condensate might be rectified by its internal rotational motion.

Recently, the dynamics of a spinor condensate with a skyrmion spin texture was theoretically investigated~\cite{HanJ,Yip,MagRelax_PRL}. In the study of a rigid spin texture case~\cite{MagRelax_PRL}, it was shown that the spin texture can precess around the center of the condensate and its precession direction is determined by the chirality of the spin texture. In terms of the relative motion between the rigid spin texture and the condensate center, the circulation direction observed in our experiment is consistent with the theoretical prediction. However, their direct comparison is limited because our system is an externally driven system.

Another remarkable observation in the condensate dynamics is the nucleation of quantized vortices. Figure 3 shows time-of-flight images of the condensate for various driving times $t_m$, where quantized vortices are detected with density-depleted cores. The condensate first deforms into an elliptical shape and after a few 100~ms, quantized vortices start to appear in its boundary region. Eventually, the condensate becomes packed with many vortices. For 2~s relaxation in a stationary trap, vortices form a triangular lattice [Figs.~3(f)-3(j)], indicating that they are of the same sign~\cite{Fetter_RMP}. For $\omega_m=2\pi\times39.2$~Hz, the vortex number $N_v$ almost linearly increases for $t_m <1$~s and becomes saturated to $N_v>100$ within 3~s [Fig.~3(k)]. There is no significant heating during the process and the condensate fraction decreases by less than 20\%~\cite{Supple}.

We attribute the vortex nucleation in the circulating condensate to the inhomogeneity of the effective magnetic field $\mathbf{B}^e(\mathbf{r})$. The effective Lorentz force $\mathbf{F}_B=\mathbf{v}_s \times \mathbf{B}^e$ has $\nabla \times \mathbf{F}_B =-(\mathbf{v}_s\cdot\nabla)\mathbf{B}^e\neq 0$ for uniform $\mathbf{v}_s$ and thus, it can play as a shearing force to deform the condensate and consequently lead to the dynamical nucleation of quantized vortices in the boundary region. We observed that the vortex nucleation is affected by the axial trapping frequency $\omega_z$~\cite{Supple}, implying that a three-dimensional description including the coupling between the axial and transverse motions might be necessary to explain the vortex nucleation dynamics. 

\begin{figure}
\includegraphics[width=7.5cm]{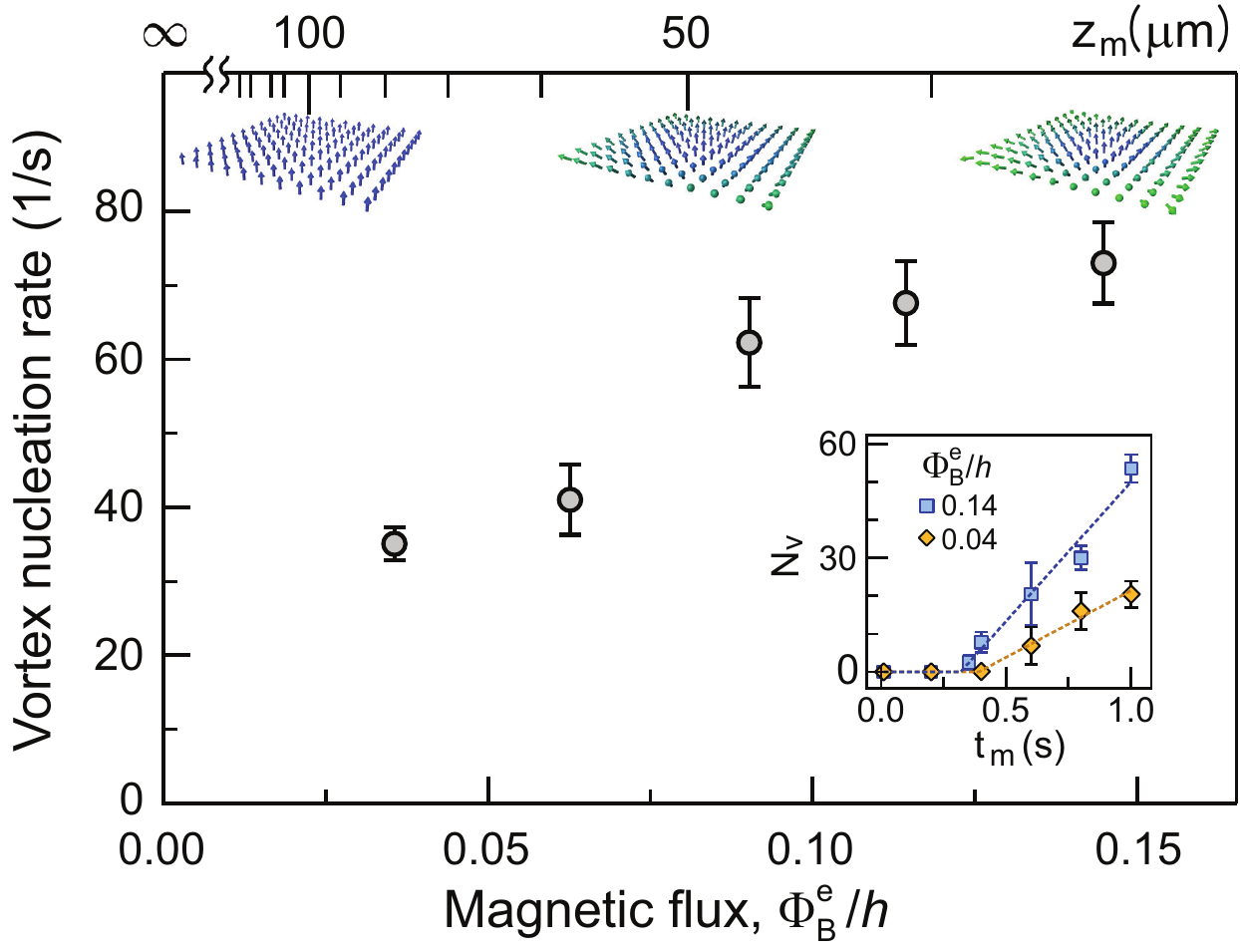}
\caption{(color online).  Vortex nucleation rate $\Gamma_v$ versus effective magnetic flux $\Phi_{B}^e$ passing through the condensate. The inset displays the vortex number $N_v$ as a function of driving time $t_m$ and $\Gamma_v$ is determined from a linear fit (dashed line) to the data. The data were acquired with condensates containing about $2.3\times10^6$ atoms for $\omega_m=2\pi\times 35.0$~Hz and $X_m=2.8~\mu$m.}
\label{Figure4}
\end{figure}

In our experiment, we observe that the nucleated vortex number shows similar asymmetric dependence on the driving frequency $\omega_m$ to that observed in the amplification rate of the $y$-directional motion~\cite{Supple}. This suggests that the vortex nucleation rate $\Gamma_v$ can be used as a quantitative measure for the geometric Hall effect in our superfluid system. We measure $\Gamma_v$ for various values of the effective magnetic flux $\Phi_{B}^e$ by changing the distance $z_m$ of the zero-field center from the condensate [Fig.~4(a)]. Here, we set $\omega_m=2\pi\times35$~Hz and adjust the radial trapping frequency $\omega_r\propto \sqrt{\frac{B_q^2}{B_z}}$ with $B_q$ and $B_z$ to obtain a maximum Hall response for a given $z_m=B_z/B_q$. The vortex nucleation rate $\Gamma_v$ was determined from a linear fit of $\Gamma_v(t_m-t_0)\times\theta(t_m-t_0)$ to $N_v(t_m)$ for $t_m\leq 1$~s [Fig.~4(a) inset], where $\theta(t)$ is the Heaviside step function. We find that $\Gamma_v$ monotonically increases with $\Phi_{B}^e$. The onset time $t_0=390\pm$50~ms, showing no clear dependence on $\Phi_{B}^e$.

In conclusion, we have observed the geometric Hall effect in a spinor Bose-Einstein condensate with a skyrmion spin texture and presented the first study of superfluid dynamics in an inhomogeneous gauge field. We expect the spin-texture oscillation method to be extended to spin mixture systems for studying spin-dependent transport phenomena such as spin drag~\cite{Spindrag} and spin Hall effects~\cite{SpinHall_Nature,HEspinor_PRB}. Employing nanofabricated ferromagnetic structures might be envisaged to achieve a strong effective magnetic field~\cite{FerroNano}.

This work was supported by the NRF of Korea funded by MSIP (Grants No. 2011-0017527, No. 2008-0061906, and No. 2013-H1A8A1003984).

\newpage
\noindent\textbf{Supplemental Material}
\vspace{0.25in}

\noindent\textbf{Experimental setup} \\
A schematic diagram of the experimental setup is depicted in Fig.~5. The optical dipole trap is formed by focusing a 1064-nm laser beam with a $1/e^2$ beam waist of 1.9~mm (17~$\mu{}m)$ in the $y$ ($z$) direction. The symmetric axis of the magnetic quadrupole field is aligned to the $z$, gravitational direction. The trapping potential for a sodium atom in the $|F=1,m_F=-1\rangle$ state can be expressed as
\begin{eqnarray}
V_t(x,y,z) &=&\frac{m}{2} (\omega_{x}^2x^2+\omega_{y}^2y^2+\omega_{z}^2z^2)  + \\
&+& \frac{1}{2} \mu_B  B_q \sqrt{\frac{(x-x_m)^2+y^2}{4}+(z-z_m)^2} + \nonumber \\
&+& mgz, \nonumber
\end{eqnarray}
where $m$ is the atomic mass, $\omega_{x,y,z}$ are the trapping frequencies of the optical potential, $\mu_B$ is the Bohr magneton, $B_q$ is the axial field gradient of the magnetic field, and $g$ is the gravitational acceleration. In the sample preparation, the magnetic levitation effect is weaker than the gravitational force, i.e. $B_q<8.1$~G/cm, allowing evaporation cooling of the sample by reducing the trap depth of the optical potential. The field gradient $B_q$ was calibrated from the free falling motion of the atoms after turning off the optical trap. The distance $z_m$ between the zero-field center and the atomic sample was determined within less than 2~$\mu$m from measurements of the atom loss rate due to the Majorana spin-flip and the resonance frequency for the $m_F=-1\rightarrow m_F=0$ transition. The radial trapping frequencies of the hybrid trap are given as $\omega_{r_x,r_y}=\sqrt{\omega_{x,y}^2+\frac{\mu_B B_q}{8m z_m}}$ in the $x$ and $y$ directions, respectively. Since the radical confinement from the magnetic field is much stronger than that from the optical potential, we estimated the radial trapping frequency as $\omega_r=\sqrt{\frac{\mu_B B_q}{8m z_m}}$ within 0.5\%. 

\newpage
\noindent\textbf{Single-particle dynamics} \\
When an atom slowly moves in the magnetic quadrupole field, it feels an effective Lorentz force $\mathbf{F}_B=\mathbf{v}\times\mathbf{B}^e$ and an effective electric force $\mathbf{F}_E=-\mathbf{v}_m \times \mathbf{B}^e$, where $\mathbf{v}=\frac{d\mathbf{r}}{dt}$ is the atom's velocity and $\mathbf{v}_m=\frac{d\mathbf{r}_m}{dt}$ is the velocity of the zero-field center of the magnetic quadrupole field [Fig.~1(c)]. Including the restoring force from the hybrid trap, the equation of motion for the atom in the $z=0$ plane is given as
\begin{eqnarray}
m \frac{d^2 \mathbf{r}_{\perp}}{d t^2}  &=&-m \omega_r^2 (\mathbf{r} - \mathbf{r}_m)_{\perp}+(\mathbf{F}_{B} +\mathbf{F}_{E})_{\perp} \\
&=&-m \omega_r^2 (\mathbf{r} - \mathbf{r}_m)_{\perp}+(\frac{d\mathbf{r}}{d t} -\frac{d\mathbf{r}_m}{d t} )_{\perp}\times {B}^e_z(r_{\perp}')\hat{z}, \nonumber
\end{eqnarray}
where $\mathbf{r}_{\perp}=(x,y)$, the subscript $\perp$ denotes the projection to the $x$-$y$ plane, and ${B}^e_z(r_{\perp}')=\frac{2\hbar z_m}{(r_\perp'^2+4z_m^2)^{3/2}}$ from Eq.~(5). We assume no $z$-directional motions of the atom and the zero-field center. In the reference frame fixed to the spin-texture, i.e. with the coordinate transformation of $\mathbf{r}'=\mathbf{r} - \mathbf{r}_m$, 
\begin{eqnarray}
m \frac{d^2 \mathbf{r}_{\perp}'}{d t^2}&=& m \frac{d^2 \mathbf{r}_\perp} {d t^2}-m \frac{d^2 \mathbf{r}_{m,\perp}}{d t^2}\\
&=&-m \omega_r^2 \mathbf{r}_{\perp}'+\frac{d \mathbf{r}_{\perp}'}{d t}\times {B}^e_z(r_{\perp}')\hat{z}-m \frac{d^2 \mathbf{r}_{m,\perp}}{d t^2}. \nonumber
\end{eqnarray}
Finally, for the spin-texture oscillations with $\mathbf{r}_m=(X_m\sin\omega_mt,0,z_m)$, we have
\begin{equation}
m \frac{d^2 \mathbf{r}_{\perp}'}{d t^2}=-m \omega_r^2 \mathbf{r}_{\perp}'+\frac{d \mathbf{r}_{\perp}'}{d t}\times {B}^e_z(r_{\perp}')\hat{z}+m\omega_m^2 X_m \sin(\omega_mt)\hat{x}.
\end{equation}

In Figures 6(a)-6(e), single-particle trajectories from Eq.~S4 are displayed for $z_m=36~\mu$m, $\omega_r=2\pi\times40.3$~Hz, $X_m=1.9~\mu$m, and various driving frequencies $\omega_m$. Figs.~6(f)-6(j) show the numerical results where we refine the restoring force term in Eq.~(10) with that from the magnetic trapping potential $2m\omega_r z_m^2\sqrt{1+\frac{r^2}{2 z_m^2}}$. The anharmonicity of the magnetic trapping potential lowers the effective trapping frequency as the oscillation amplitude increases, and might explain the negative shift of the driving frequency $\omega_{m}$ from $\omega_r$ for the maximum Hall response, which becomes noticeable with smaller $z_m$ [Fig.~8(e)].

\newpage
\begin{figure*}
\includegraphics[width=8.0cm]{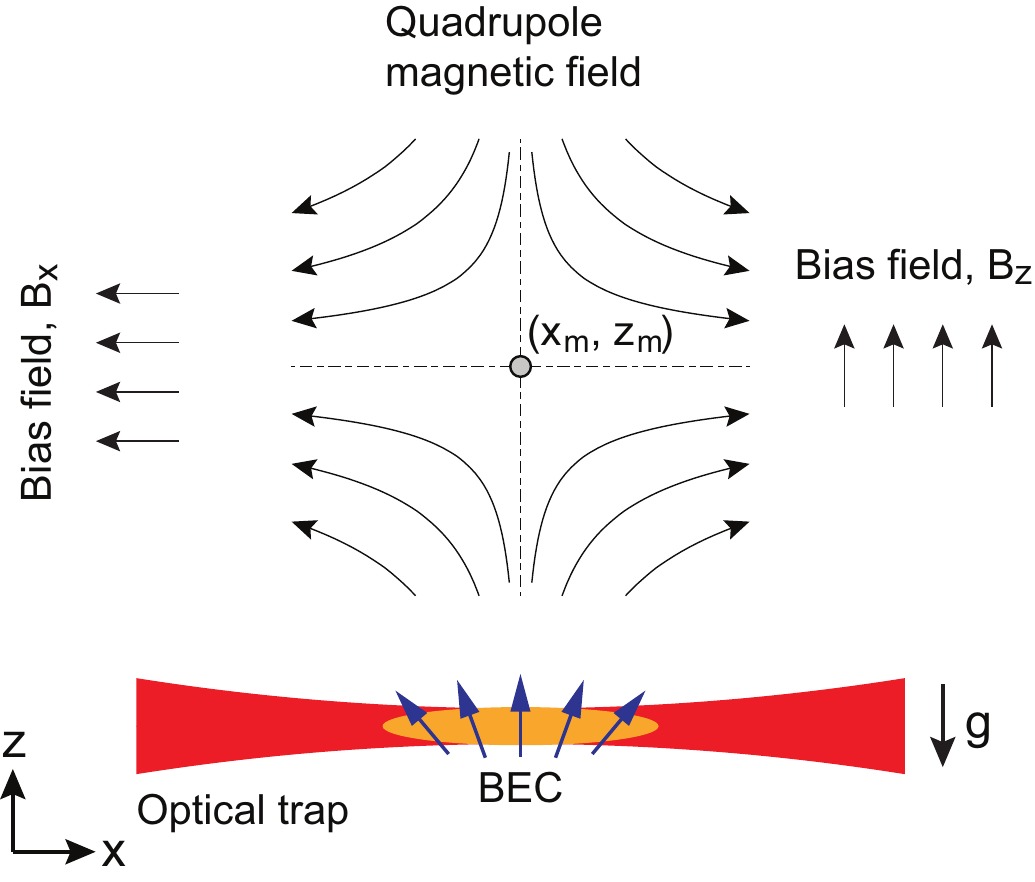}
\caption{Schematic of the experimental setup. A spinor condensate of sodium atoms is confined in a hybrid trap formed by a focused laser beam and a magnetic quadrupole field. The relative position of the quadrupole field to the optical trap is controlled by the bias fields. The blue arrows indicate the local magnetic field direction in the condensate.}
\label{FigS1}
\end{figure*}

\begin{figure*}[b]
\includegraphics[width=14.5cm]{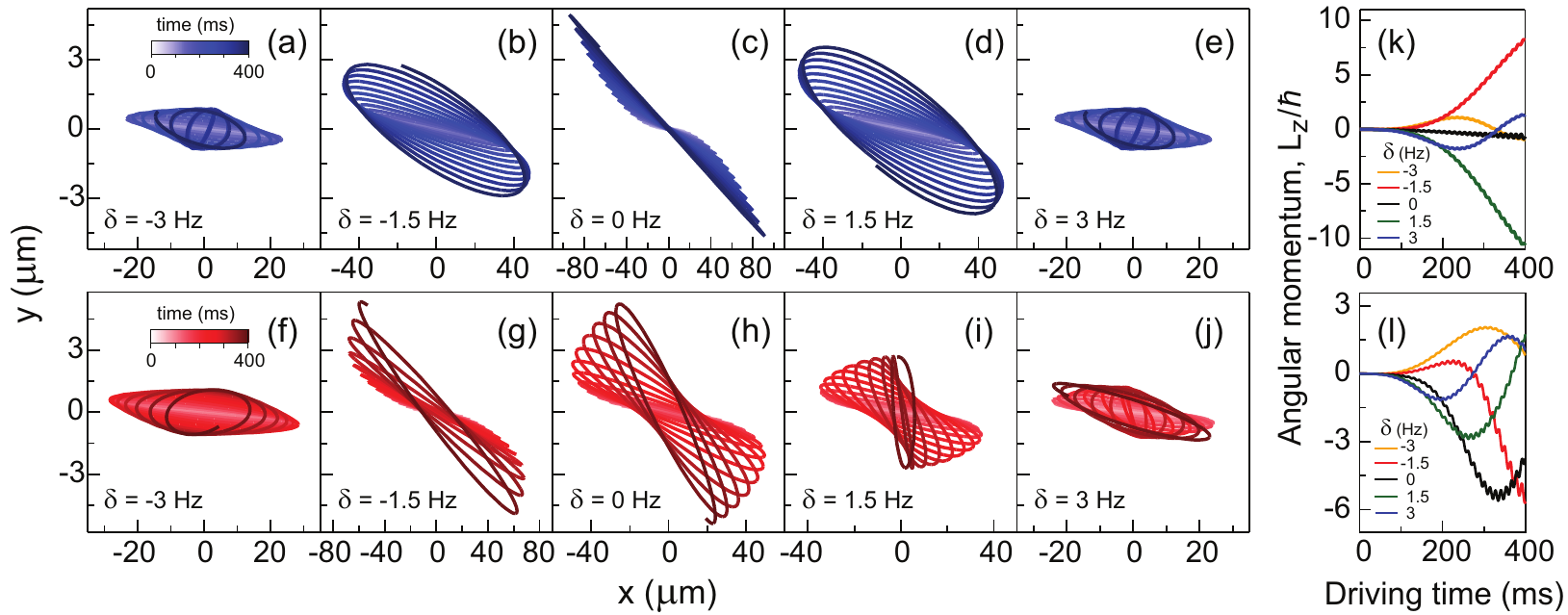}
\caption{Single-particle trajectory under the spin-texture oscillations. (a)-(e) Numerical results obtained from Eq.~(10) for $z_m=36~\mu$m, $\omega_r=2\pi\times40.3$~Hz, $X_m=1.9~\mu$m, and various detunings of the driving frequency $\delta=(\omega_m-\omega_r)/2\pi$. (f)-(j) Numerical results taking into account the anharmonicity of the magnetic trapping potential. (k) and (l) show the temporal evolution of the angular momentum of the particle with respect to the trap center for (a)-(e) and (f)-(j), respectively.}
\label{FigS2}
\end{figure*}

\begin{figure*}
\includegraphics[width=13.0cm]{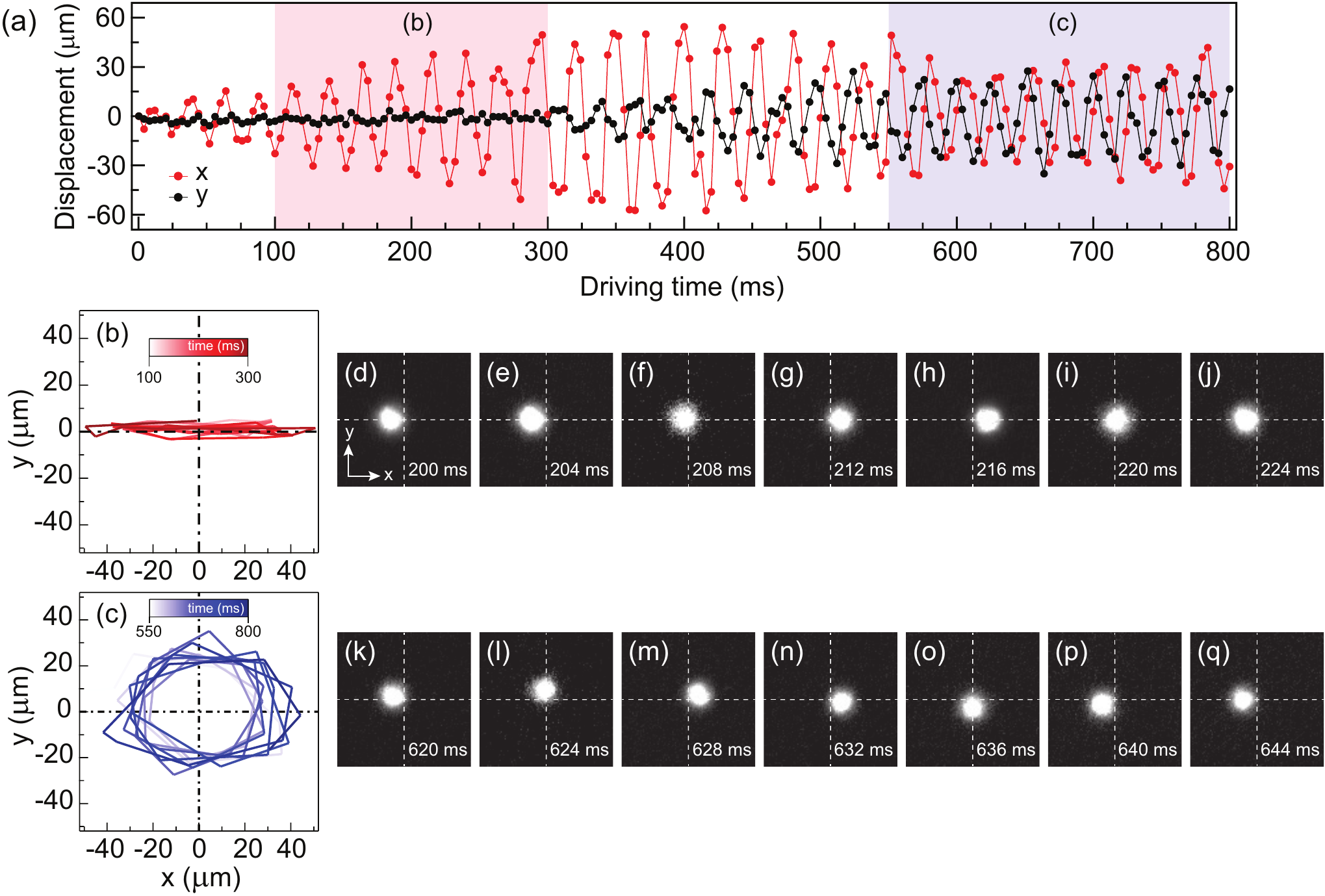}
\caption{Temporal evolution of the condensate position under the spin-texture oscillations. (a) the same data in Fig.~2(b) with $\omega_m=2\pi\times39.2$~Hz. Trajectory of the condensate in the $x$-$y$ plane for (b) driving time $t_m=100\sim300$~ms [red zone in (a)] and (c) $t_m=550\sim800$~ms [blue zone in (a)]. The initial linear motion evolves into a circular motion. (d)-(j) and (k)-(q) show in-trap images of the condensate for one oscillation period in (b) and (c), respectively. The field of view in the images is $300~\mu$m$\times 300~\mu$m. We applied a 100-$\mu$s pulse of a resonant light just before taking an in-trap absorption image to reduce the peak optical density of the sample to less than 5, and determined the center position of the condensate from a 2D Gaussian fit to the image.}
\label{FigS3}
\end{figure*}

\begin{figure*}
\includegraphics[width=14cm]{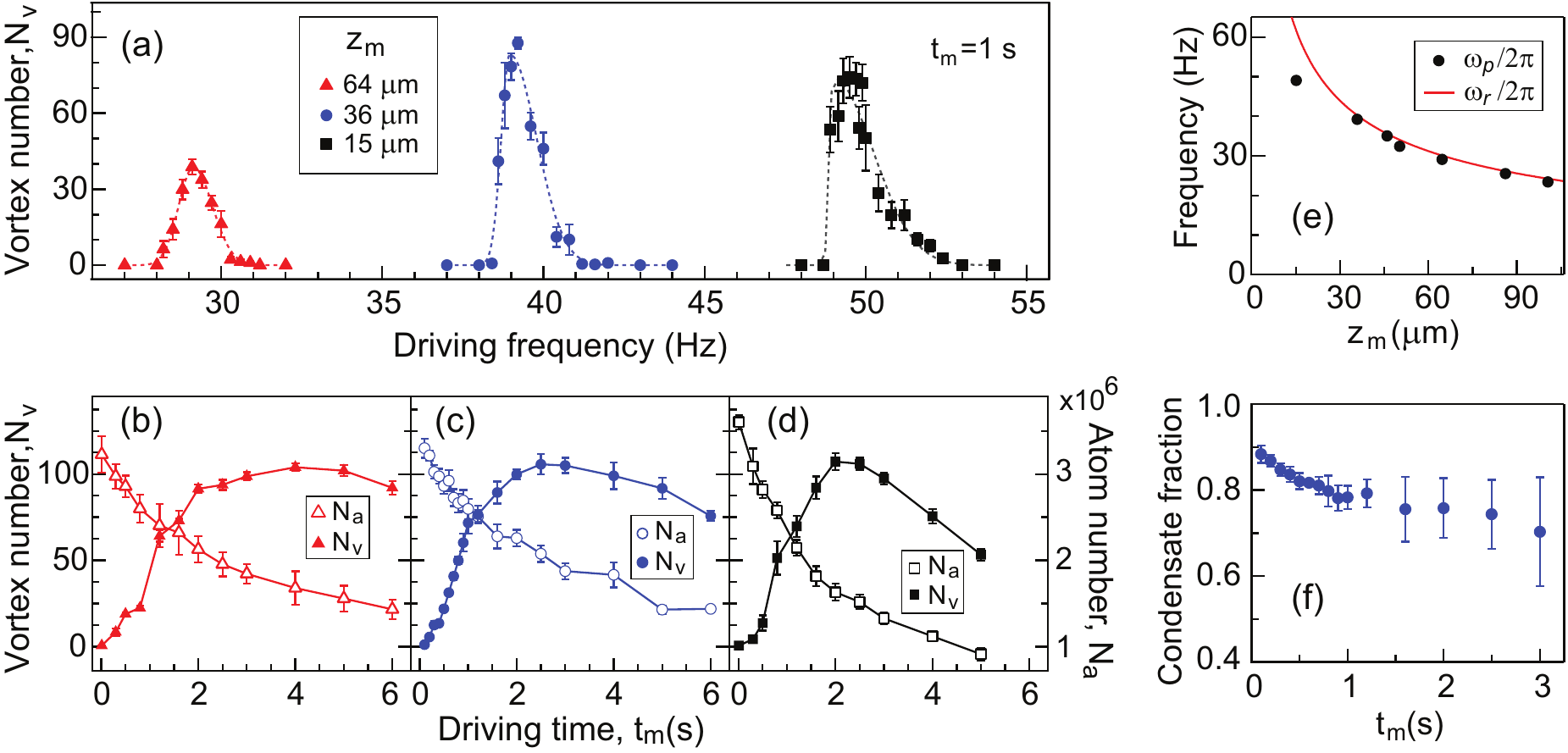}
\caption{(a) Number of nucleated vortices for $t_m=1$~s as a function of the driving frequency $\omega_m/2\pi$. The distance $z_m$ of the zero-field center to the condensate was changed with $B_z$ for $B_q=7.6$~G/cm. The vortex number $N_v$ was measured after moving the zero-field center to $z_m=102~\mu{}m$ in 1~s and a subsequent 1-s holding. Dashed lines are asymmetric Gaussian fits to the vortex number. The peak-response driving frequency, and the left and right $1/e^2$ linewidths are $(\omega_p,\gamma_L,\gamma_R)/2\pi = (29.1, 0.6, 0.8)$~Hz for $z_m=64~\mu$m, $(39.2, 0.4, 1.2)$~Hz for $z_m=36~\mu$m, and $(49.0, 0.2, 1.8)$~Hz for $z_m=15~\mu$m. Temporal evolution of the vortex number $N_v$ (solid) and the atom number $N_a$ (open) for (b) $(z_m,\omega_m/2\pi)=(64~\mu$m, 29.1~Hz), (c) (36~$\mu$m, 39.2~Hz), and (d) (15~$\mu$m, 49~Hz). Each data point was obtained from at least 8 measurements and the error bars represent the standard deviation of the measurements. (e) The peak-response driving frequency $\omega_p$ (solid circle) and the estimated trapping frequency $\omega_r$ (red line) as a function of $z_m$. (f) The condensate fraction as a function of $t_m$ for $z_m=36~\mu$m. The condensate fraction was determined from a bimodal fit to an image like Fig.~3(f)-(j).}
\label{FigS4}
\end{figure*}

\begin{figure*}
\includegraphics[width=8.5cm]{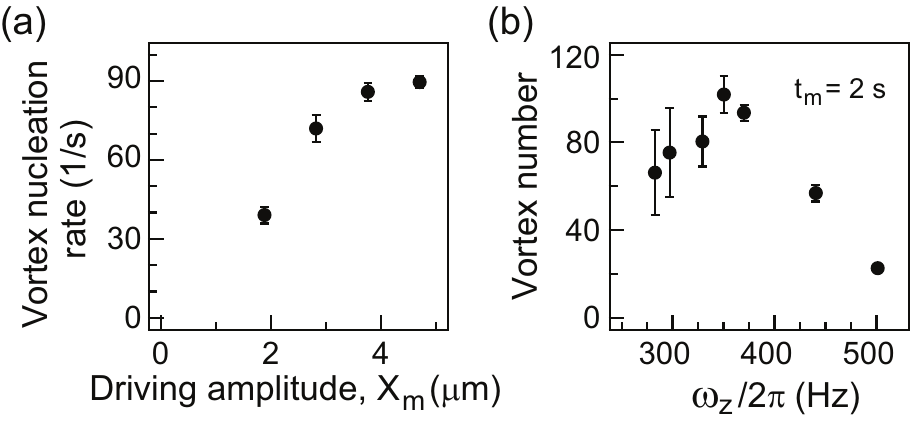}
\caption{(a) Vortex nucleation rate versus driving amplitude $X_m$ ($B_q=7.6$~G/cm, $z_m=36~\mu$m, and $\omega_m=2\pi\times39.2$~Hz). The vortex nucleation rate was determined from a linear fit to the vortex number for $t_m\leq 1$~s (see the inset of Fig.~4). (b) Vortex number $N_v$ as a function of the axial trapping frequency $\omega_z$ ($B_q=7.6$~G/cm, $z_m=36~\mu$m, $X_m=1.9~\mu$m, $\omega_m=2\pi\times39.2$~Hz, and $t_m=2$~s). Each data point is obtained from at least six measurements.}
\label{FigS5}
\end{figure*}

\begin{figure*}
\includegraphics[width=10cm]{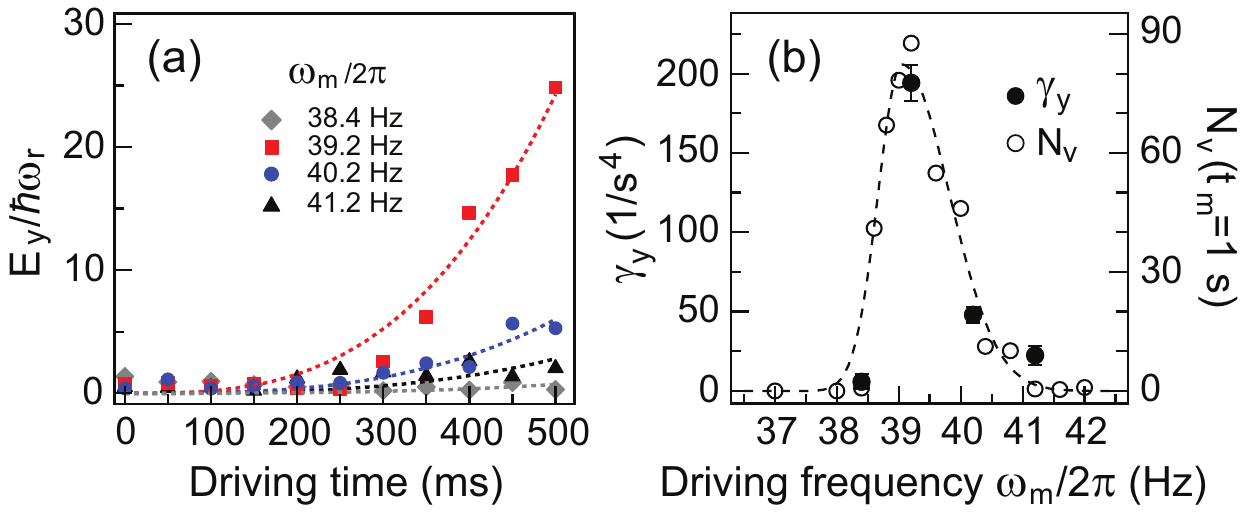}
\caption{Temporal evolution of the $y$-directional mechanical energy $E_y=m\omega_r^2 \langle y^2\rangle$ for the data in Figs.~2(a)-2(d). $\langle y^2\rangle$ was obtained by averaging over a 50-ms time bin. In the single particle dynamics of Eq.~(10), $\langle y^2\rangle \propto  t^4$ on resonance with $\omega_m=\omega_r$. The dashed lines are curve fits of $E_y/\hbar\omega_r=\gamma_y t_m^4$ to the data points of each $\omega_m$. (b) $\gamma_y$ versus $\omega_m/2\pi$, displayed together with the vortex number data from Fig.~3(l).}
\label{FigS6}
\end{figure*}

\begin{figure*}
\includegraphics[width=11cm]{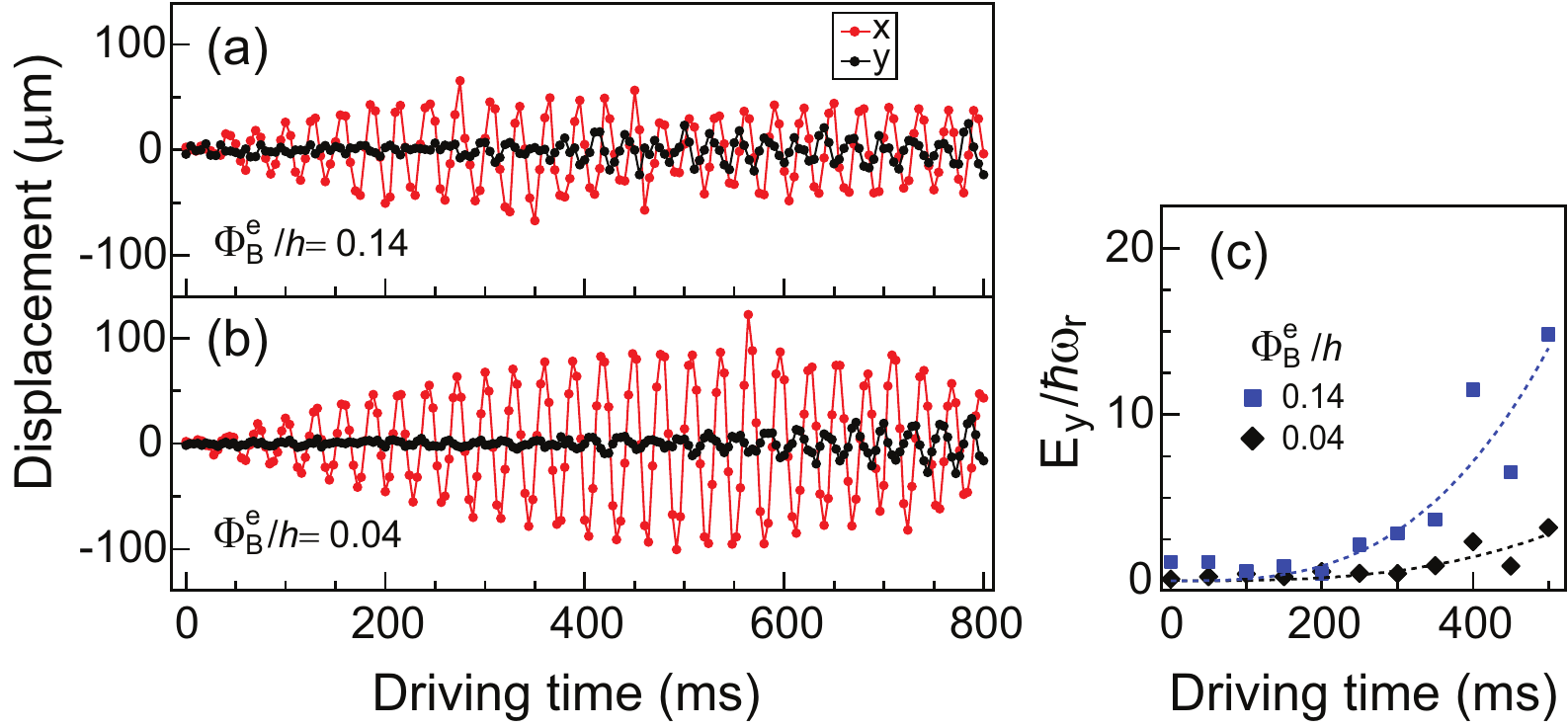}
\caption{Temporal evolution of the condensate position for (a) $\Phi_B^e/h=0.14$ and (b) 0.04 in Fig.~4, and (c) the corresponding evolution of the $y$-directional mechanical energy $E_y(t_m)$. The dashed lines are the curve fits of $E_y/\hbar\omega_r=\gamma_y t_m^4$ to the data.}
\label{FigS7}
\end{figure*}


\begin{references}
\bibitem{footnote2} In order to reduce the optical density of the sample, we applied a 100-$\mu$s pulse of a resonant light to the sample just before imaging. The modulation amplitude $X_m$ was checked to be in the linear-response regime in terms of vortex nucleation~\cite{Supple}.
\end{references}
\end{document}